\documentclass[twocolumn]{aastex61}

\usepackage{epsfig}
\usepackage{epstopdf}
\usepackage{amsmath}
\usepackage{amssymb}
\usepackage{subfigure}
\usepackage{float}
\usepackage{ulem}  
\usepackage{url,hyperref}
\usepackage{footmisc}
\usepackage{tabularx}
\usepackage{natbib}

\newcommand{\hess}{H.E.S.S.~}
\newcommand{\g}{$\gamma$}
\newcommand{\gs}{$\gamma \ $}
\newcommand{\mrk}{Mrk~501~}

\shorttitle{Mrk 501 2014 flare and LIV}
\shortauthors{H.E.S.S. Collaboration}
\begin{document}

\title{The 2014 TeV \g-ray flare of Mrk 501 seen with H.E.S.S.: temporal and spectral constraints on Lorentz invariance violation}

\AuthorCallLimit=300
\collaboration{(H.E.S.S. Collaboration)}\altaffiliation{corresponding author (contact.hess@hess-experiment.eu)}
\noaffiliation\email{contact.hess@hess-experiment.eu}
\author{H.~Abdalla}\affiliation{Centre for Space Research, North-West University, Potchefstroom 2520, South Africa}
\author{F.~Aharonian}\affiliation{Max-Planck-Institut f\"ur Kernphysik, P.O. Box 103980, D 69029 Heidelberg, Germany}\affiliation{Dublin Institute for Advanced Studies, 31 Fitzwilliam Place, Dublin 2, Ireland}\affiliation{National Academy of Sciences of the Republic of Armenia,  Marshall Baghramian Avenue, 24, 0019 Yerevan, Republic of Armenia}
\author{F.~Ait~Benkhali}\affiliation{Max-Planck-Institut f\"ur Kernphysik, P.O. Box 103980, D 69029 Heidelberg, Germany}
\author{E.O.~Ang\"uner}\affiliation{Aix Marseille Universit\'e, CNRS/IN2P3, CPPM, Marseille, France}
\author{M.~Arakawa}\affiliation{Department of Physics, Rikkyo University, 3-34-1 Nishi-Ikebukuro, Toshima-ku, Tokyo 171-8501, Japan}
\author{C.~Arcaro} \affiliation{Centre for Space Research, North-West University, Potchefstroom 2520, South Africa}
\author{C.~Armand} \affiliation{Laboratoire d'Annecy-le-Vieux de Physique des Particules, Universit\'{e} Savoie Mont-Blanc, CNRS/IN2P3, F-74941 Annecy-le-Vieux, France}
\author{M.~Arrieta}\affiliation{LUTH, Observatoire de Paris, PSL Research University, CNRS, Universit\'e Paris Diderot, 5 Place Jules Janssen, 92190 Meudon, France}
\author{M.~Backes}\affiliation{University of Namibia, Department of Physics, Private Bag 13301, Windhoek, Namibia}\affiliation{Centre for Space Research, North-West University, Potchefstroom 2520, South Africa}
\author{M.~Barnard}\affiliation{Centre for Space Research, North-West University, Potchefstroom 2520, South Africa}
% ...

%\altaffiliation{Creator of AASTeX v6.1}
%\collaboration{(LaTeX collaboration)}
%\affiliation{AAS Director of Publishing}
%\affiliation{American Astronomical Society \\
%2000 Florida Ave., NW, Suite 300 \\
%Washington, DC 20009-1231, USA}



\begin{abstract}
The blazar Mrk~501 ($z=0.034$) was observed at very-high-energy gamma rays (VHE, $E\gtrsim 100$~GeV) during a bright flare on the night of June 23-24 2014 (MJD 56832) with the H.E.S.S. \mbox{phase-II} array of Cherenkov telescopes. Data taken that night by H.E.S.S. at large zenith angle reveal an exceptional number of gamma-ray photons at multi-TeV energies, with rapid flux variability and an energy coverage extending significantly up to 20 TeV. This data set is used to constrain Lorentz invariance violation (LIV) using two independent channels: a temporal approach considers the possibility of an energy dependence in the arrival time of gamma rays, whereas a spectral approach considers the possibility of modifications to the interaction of VHE gamma rays with extragalactic background light (EBL) photons. The non-detection of energy-dependent time delays and the non-observation of deviations between the measured spectrum and that of a supposed power-law intrinsic spectrum with standard EBL attenuation are used independently to derive strong constraints on the energy scale of LIV ($E_{\rm{QG}}$) in the subluminal scenario for linear and quadratic perturbations in the dispersion relation of photons. For the case of linear perturbations, the 95\% confidence level limits obtained are $E_{\rm{QG},1}>3.6 \times 10^{17} \ \rm{GeV}$ using the temporal approach and $E_{\rm{QG},1}>2.6 \times 10^{19} \ \rm{GeV}$ using the spectral approach. For the case of quadratic perturbations, the limits obtained are  $E_{\rm{QG},2}> 8.5 \times 10^{10} \ \rm{GeV}$ using the temporal approach and $E_{\rm{QG},2}> 7.8 \times 10^{11} \rm{ GeV}$ using the spectral approach.
\end{abstract}




\section{Introduction}
\label{Section:Intro}
Blazars are commonly considered to be active galactic nuclei with their jets closely aligned with the line of sight to the observer \citep{1995PASP..107..803U}. They exhibit flux variability on time-scales ranging from years to minutes over the entire electromagnetic spectrum, from radio to very high energy (VHE, $E \gtrsim 100$~GeV) $\gamma$ rays. The observation of flaring activity of blazars at VHE provides insights into the acceleration mechanisms involved at the source. These observations are also relevant for the study of propagation effects not directly related to the source. This includes fundamental physics aspects like Lorentz invariance violation (LIV).

Lorentz invariance has been established to be exact up to the precision of current experiments.
Some approaches to quantum gravity (QG) suggest, however, that Lorentz symmetry could be broken at an energy scale thought to be around the Planck scale $ \left( \text{E}_{\text{Planck}} = \sqrt{\hbar c^5 /G} \simeq 1.22 \times 10^{19} \text{~GeV} \right)$, see \textit{e.g.} \citet{Jacobson:2005bg,AmelinoCamelia:2008qg,Mavromatos:2010pk}.
A generic approach to LIV effects for photons consists in adding an extra term in their energy-momentum dispersion relation:
\begin{equation}
E^2 \simeq p^2c^2 \left[ 1 \pm \left(\frac{E}{E_{\text{QG}}} \right)^n \ \right],
\label{Eq:DispersionLIV}
\end{equation}
where $E$ and $p$ are the energy and momentum of the photon, $E_{\text{QG}}$ is the hypothetical energy scale at which Lorentz symmetry would be broken, and $n$ is the leading order of the LIV perturbation. The sign of this perturbation is model-dependent and refers to subluminal ($-$) and superluminal ($+$) scenarios. In some theoretical models the sign of the perturbation can also be related to the polarization of the particle.

A non-infinite value of $E_{\text{QG}}$ in Eq.~\ref{Eq:DispersionLIV} would induce non-negligible observational effects. It would cause an energy-dependent velocity of photons in vacuum which in turn would translate into an energy-dependent time-delay in the arrival time of $\gamma$ rays traveling over astrophysical distances \citep{Amelino:1998,Ellis:2011ek}. Another interesting effect is on the kinematics of photon interactions like the production of electron-positron pairs from the interaction of VHE $\gamma$ rays with photons of the extragalactic background light (EBL), resulting in deviations with respect to standard EBL attenuation in the energy spectrum of blazars \citep{Stecker:2001vb,jacob2008inspecting}.\\

Valuable constraints on $E_{\text{QG}}$ considering linear ($n=1$) or quadratic ($n=2$) perturbations in Eq.~\ref{Eq:DispersionLIV} have already been obtained from the observations of several $\gamma$-ray bursts (GRBs) at high energy (HE, 100~MeV $\lesssim E \lesssim$ 100~GeV) and flares of blazars at VHE, mostly looking for energy-dependent time delays (for a review see \textit{e.g.} \citet{Horns:2016} and references therein). With H.E.S.S., temporal LIV studies have in particular been conducted using flares of the blazars PKS~2155-304 ($z=0.116$) \citep{HESS:2011aa} and PG~1553+113 ($z\simeq 0.49$) \citep{Abramowski:2015}. For the linear case, the best existing limits are obtained using GRBs and have reached the Planck scale \citep{Vasileiou:2013}. The constraints on the quadratic term remain several orders of magnitude below the Planck scale and will continue to be a challenge for future studies. % as on the contrary to the linear corrections they do not violate CPT symmetry

Both the temporal and spectral LIV effects can be used to put competitive constraints on $E_{\text{QG}}$ using VHE \g-ray observations of a blazar flare, given certain conditions on the energy coverage and distance to the source.\\

Markarian 501 (Mrk 501) is a well-known nearby blazar at a redshift $z = 0.034$ \citep{Mkn501Redshift}. It was the second extragalactic source discovered at VHE in 1995~\citep{Quinn:1996dj} and has been extensively monitored since then. In 1997, Mrk~501 showed an exceptional flare at VHE with an integral flux up to four times the flux of the Crab Nebula \citep{1997ApJ...487L.143C, 2000ApJ...536..742P, 1999A&A...349...11A, 1999A&A...350...17D}. The hard VHE spectrum extending up to $\sim20$~TeV measured by HEGRA \citep{Aharonian:1999vy,Aharonian:2000xr} during this 1997 flare triggered a wide interest on EBL attenuation and LIV (see \textit{e.g.} \citealt{Aharonian:2001cp,Tavecchio:2015rfa}). In 2005, rapid flux variations observed at VHE by MAGIC \citep{Albert:2007zd} also triggered interest for LIV from the point of view of energy-dependent time delays \citep{Albert:2007qk}. 

In 2014, the monitoring\footnote{\url{http://fact-project.org/monitoring}} of Mrk 501 with the First G-APD Cherenkov Telescope (FACT) \citep{2013JInst...8P6008A,Biland:2014fqa, Dorner:2015jka} led to the detection of several high-state events which triggered observations with the H.E.S.S. experiment. On the night of June 23-24 2014 (MJD 56832) a flare comparable to the 1997 maximum was observed with the full array of \hess telescopes. This flare corresponds to the highest flux level of Mrk~501 ever recorded with the \hess telescopes. Data analysis reveals an exceptional $\gamma$-ray flux at multi-TeV energies, with a rapid flux variability and an energy spectrum extending up to 20~TeV. This data set thus has excellent properties for the investigation of LIV effects through both temporal and spectral channels. 

This paper is organized as follows. The H.E.S.S. observations of the 2014 flare of \mrk and the data analysis are described in Sec.~\ref{Section:HESS}. The temporal study of the flare is presented in Sec.~\ref{Section:Temp}, focusing on the search for LIV with time delays. The spectral study of the flare is presented in Sec.~\ref{Section:Spec}, investigating the possibility of LIV through modifications to standard EBL attenuation. Results are discussed and summarized in Sec.~\ref{Section:Summary}.

\section{\hess observations and data analysis}
\label{Section:HESS}

H.E.S.S. is an array of five imaging atmospheric Cherenkov telescopes located in the Khomas Highland, Namibia ($23^\circ 16'18''$ S, $16^\circ 30'01''$ E), at an elevation of $1800\,$m above sea level. H.E.S.S. is the first hybrid array of Cherenkov telescopes since the addition in 2012 of a fifth $28\,$m diameter telescope (CT5) at the center of the original array of four $12\,$m diameter telescopes (CT1-4). This configuration (\hess \mbox{phase-II}) can trigger on events detected either by CT5 alone (monoscopic events), or by any combination of two or more telescopes (stereoscopic events). Reconstruction and analysis can be performed in different modes depending on the selection of monoscopic and stereoscopic events. To fully exploit all the available information, a combined mode makes use of both monoscopic and stereoscopic events. In case of an event for which both monoscopic and stereoscopic reconstructions are possible, the choice is made depending on the uncertainty on the reconstructed direction \citep{Holler:2015tca,Holler:2015uca}.

The H.E.S.S. observations of \mrk over the month of June 2014 have been reported in \citet{Cologna:2016cws}. The presented work only regards \hess data taken on MJD 56832. Four consecutive observation runs ($\sim$28 min each) were taken on \mrk that night, with the participation of all five telescopes. These four runs pass the standard \hess data-quality selection criteria \citep{Aharonian:2006pe}, yielding an exposure of 1.8 h live time. \mrk being a northern-sky blazar, \hess observations were taken at large zenith angles, between $63^{\circ}$ and $65^{\circ}$. At such large zenith angles, both the increased atmospheric absorption as well as the increased size of the Cherenkov light pool lead to a reduced Cherenkov light density at the ground. This causes the energy threshold to be particularly high ($\gtrsim 1$~TeV). On the other hand, the effective area is enhanced at the highest energies due to the increased geometrical area covered by the light pool of inclined showers \citep{Aharonian:2005ib}.

Data reconstruction is performed using the \textit{Model Analysis} technique \citep{deNaurois:2009ud} in which recorded air-shower images are compared to template images pre-calculated using a semi-analytical model and a log-likelihood optimization technique. The combined analysis mode taking into account CT5 monoscopic, CT1-5 stereoscopic and CT1-4 stereoscopic events is used for an optimal energy coverage. A selection criterion on the image charge of 60 photo-electrons is applied. The on-source events are taken from a circular region centered around \mrk with a radius of $0.1225 \degr$. This relaxed cut on the aperture is motivated by the large signal over background ratio. The background is estimated using the \textit{Reflected Region} method described in \citet{Berge:2006ae}.

In the signal region 1930 events are observed, versus 334 events in the background region. With a solid angle ratio of 8.95 between the background and signal regions, this translates into a signal over background ratio of 46.5 and an excess of 1889.3 \gs rays detected with a significance of 83.3$\sigma$, following the statistical approach of \citet{1983ApJ...272..317L}. 
Two cross-check analyses based on a different calibration chain yield compatible results. The first follows an adaptation of the method described in \citet{Aharonian:2006pe} to allow the analysis of CT1-5 stereoscopic events and the second is based on the analysis of CT5 monoscopic events as described in \citet{2015arXiv150900794M} \footnote{At the time of writing, these cross-check analyses had no combined analysis capability.}.
 
\section{Temporal study}
\label{Section:Temp}

\subsection{Rapid flux variability}

\hess observations of this flare show rapid flux variations at multi-TeV energies. Earlier observations of \mrk at VHE have shown variations down to timescales of a few minutes \citep{Albert:2007zd}. However, these previously-reported flares were dominated by photons of energies of a few hundred GeV. Because of the large zenith angle observations with H.E.S.S., the variability observed during this flare is restricted to TeV energies.
The average integral flux above 1~TeV observed from \mrk during the peak of this flare is $I(> \text{1 TeV}) = \left( 4.4\pm 0.8_{\text{stat}} \pm 1.8_{\text{sys}} \right) \times 10^{-11} \rm{cm}^{-2}\rm{ s}^{-1}$.
There is evidence for multi-TeV flux variations on time-scales of minutes. The atmospheric transparency is verified to be stable over the course of observations using the transparency coefficient described in \citet{2014APh....54...25H}, therefore no significant spurious variability can be attributed to variations of the Cherenkov light yield (\textit{e.g.} due to clouds).

This flare shows an excess variance, as defined in \citet{2003MNRAS.345.1271V}, of $F_{\rm{var}} = 0.188 \pm 0.003$, for a time binning of seven minutes.
Considering a longer time window capturing the rise and fall of the flare, an even larger value, $F_{\rm{var}} = 1.03\pm0.01$ is obtained.  
The detailed discussion on astrophysical implications of this rapid variability relative to the long-term activity of \mrk seen in \gs rays by \hess along with FACT and Fermi-LAT is left for a dedicated forthcoming paper.

\subsection{LIV: time of flight study}
\label{Section:TOF}

The rapid flux variability at multi-TeV energies observed during the flare of Mrk~501 is used to constrain the LIV scale ($E_{QG}$) through the search for energy-dependent time delays as outlined in Sec.~\ref{Section:Intro}.
Assuming the LIV-modified dispersion relation of Eq.~\ref{Eq:DispersionLIV}, the relative energy-dependent time delay due to LIV effects for two photons with an energy difference $\Delta E_{n} = E_1^n - E_2^n$ and a time difference $\Delta t_n$ can be expressed as in \citet{Jacob2008}:

\begin{equation}
\label{eq:timez5}
\tau_n = \frac{\Delta t_n}{\Delta E_{n}} \simeq \pm \frac{n+1}{2}\,\frac{1}{E_{QG}^n} \int_0^z \frac{(1+z')^n}{H(z')} \mathrm{d}z',
\end{equation}

where $H(z) = H_0 \sqrt{\Omega_m\,(1+z)^3 + \Omega_\Lambda}$, assuming a flat $\Lambda$CDM cosmology with Hubble constant $H_0 = 67.74\ \mathrm{km \cdot Mpc^{-1}\cdot s^{-1}}$, matter density parameter $\Omega_m = 0.31$ and dark energy density parameter $\Omega_{\Lambda} = 0.69$ \citep{Ade:2015xua}. In the following, $\tau_n$ values are estimated using a likelihood method.

\subsubsection{Likelihood method}

The maximum likelihood (ML) method for the extraction of energy-dependent time-lags was first proposed in \citet{Martinez:2009} and then extensively applied for LIV analyses in H.E.S.S. with the flares of PKS~2155-304 \citep{HESS:2011aa} and PG~1553+113 \citep{Abramowski:2015}. The ML method relies on the definition of a probability density function (PDF) that describes the probability of observing a photon at energy $E$ and arrival time $t$, assuming an energy-dependent delay function $D(E_s,\tau_n)$, where $E_s$ is the energy at the source. As the data shows a very high signal over background ratio (46.5), the background contribution is neglected for the PDF. For each event, the PDF can be written as proposed in \citet{Martinez:2009}:

\begin{equation}
\begin{split}
\frac{\mathrm{d}P}{\mathrm{d}E\mathrm{d}t} = \frac{1}{N(\tau_n)} \int_0^{\infty} \Gamma({E_s}) C(E_s,t) G \left[ E,E_s,\sigma(E_s )\right] \\
F_s \left[ t - D(E_s,\tau_n) \right] \mathrm{d}E_s
\end{split}
\label{eq:PDF}
\end{equation}
where $N(\tau_n)$ is a normalization factor, $\Gamma(E_s)$ is the photon energy distribution at the source, $C(E_s, t)$ is the collection area and $G \left[ E,E_s,\sigma(E_s )\right]$ is the instrument energy response function. $F_s(t_s)$ is the emission-time distribution at the source, \textit{i.e.} without any LIV time-delay. In previous LIV studies with H.E.S.S \citep{HESS:2011aa, Abramowski:2015}, the template $F_s(t)$ was estimated from low energy events (below an energy $E_{cut}$), assuming no LIV time-lag (\textit{i.e.} $D(E_s,\tau_n) = \tau_n E^n$). In the present analysis, due to the high threshold ($\gtrsim 1$~TeV), LIV time-lag effects on the template are taken into account  and $D$ is defined as $ D(E_s,\tau_n) = \tau_n E^n - \tau_n \overline{E_T}^n$ where $\overline{E_T}$ is the mean energy of the events in the template energy range. The likelihood is a function of parameter $\tau_n$, and is built using a selection of events above $E_{cut}$, multiplying their PDF together:

\begin{equation}
L(\tau_n) = \prod_{i}P_i(t_i,E_i,\tau_n).
\label{eq:Likelihood}
\end{equation}

\subsubsection{Data selection}
\begin{figure}
\includegraphics[width=1\linewidth]{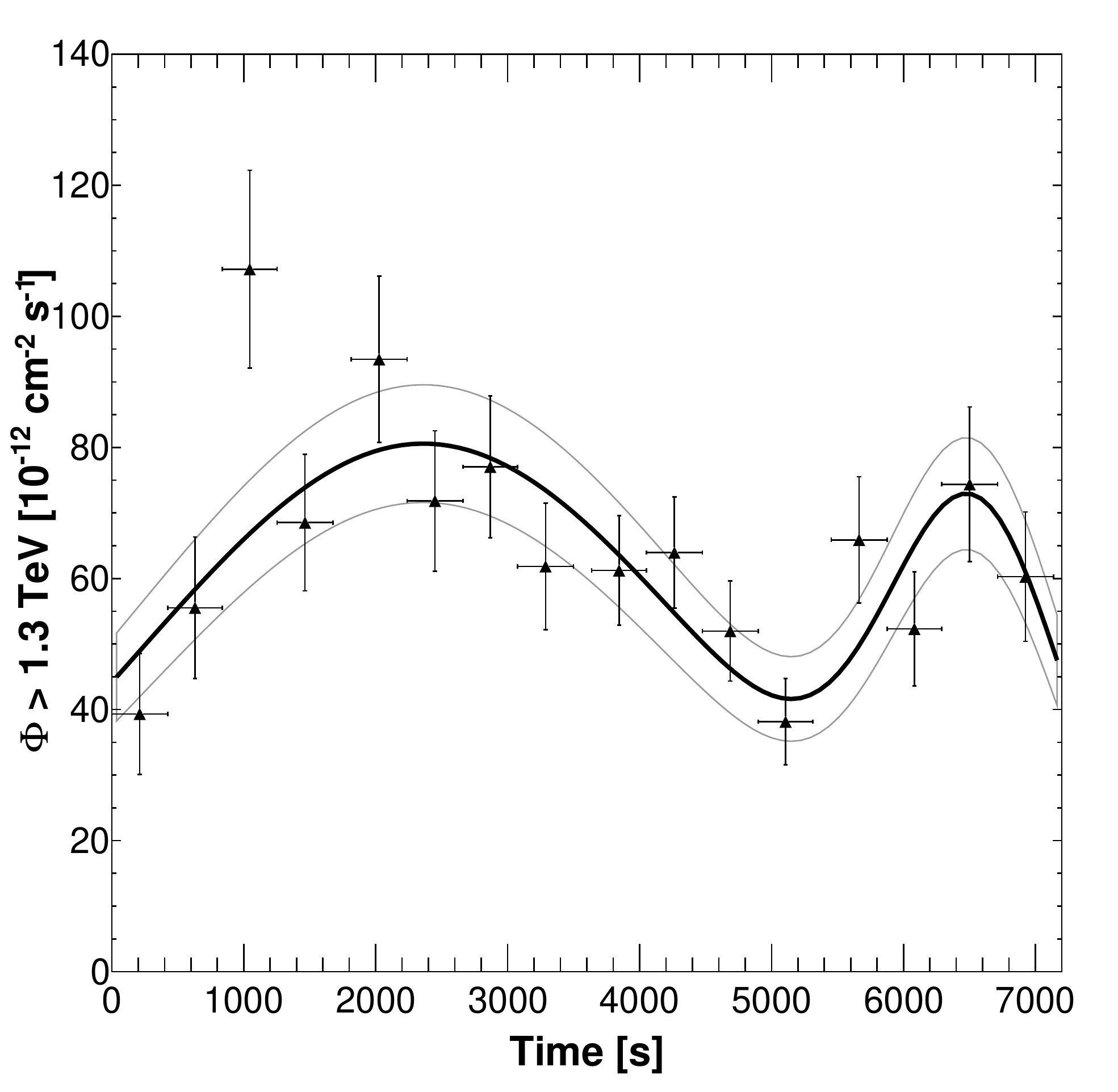}
\caption{Light curve used for $F_s(t)$ estimation in the range 1.3~$<$~E~$<$~3.25~TeV. The thick line corresponds to the best fit and the thin ones to the $1\sigma$ error envelope. The parameters of the fit function are shown in Table~\ref{temp_par}.}
\label{template_lc}
\end{figure}

From the full data sample described in Sec.~\ref{Section:HESS}, two regions are defined with two energy selections. At low energies, the template region is defined for which 1.3~$<$~E~$<$~$E_{cut}$~=~$3.25\,$TeV. The threshold value of $1.3\,$TeV corresponds to the energy at which the effective area of these observations reaches $15\%$ of its maximum value. The 773 events in the template range are used to estimate the function $F_s(t)$ by fitting their time-distribution. The template fit is shown on Fig.~\ref{template_lc} and chosen as the sum of two Gaussian functions. The result of the fit yields a $\chi^2/ndf$ of 15.9/10. The double Gaussian function is favored over a Gaussian for which $\chi^2/ndf =  38.1/13$. The fit parameters and associated errors are given in Table~\ref{temp_par}. The 662 events above $3.25\,$TeV are used to compute the likelihood and obtain the best estimate $\tau_{n, \text{best}}$. The energy cut at $3.25\,$TeV is chosen as a trade off between a robust estimation of $F_s(t)$ and the largest number of events for the likelihood calculation. The photon energy distribution $\Gamma(E_s)$ is obtained from a power law fit approximation above $E_{cut}$ with a resulting index of $3.1 \pm 0.1$.

\subsubsection{Results}
\begin{figure*}
\begin{center}
\includegraphics[scale=0.3]{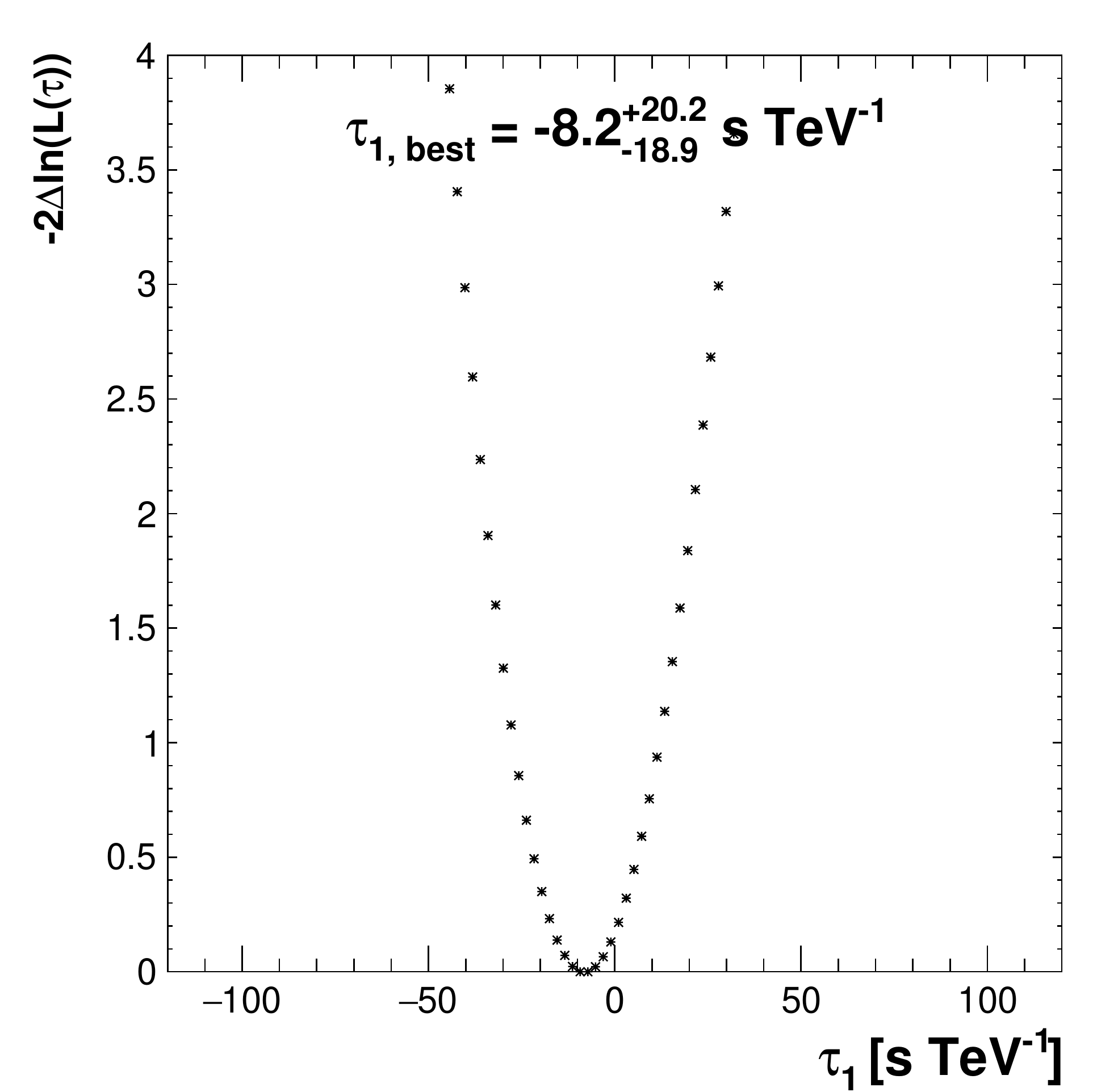}
\bigskip
\includegraphics[scale=0.3]{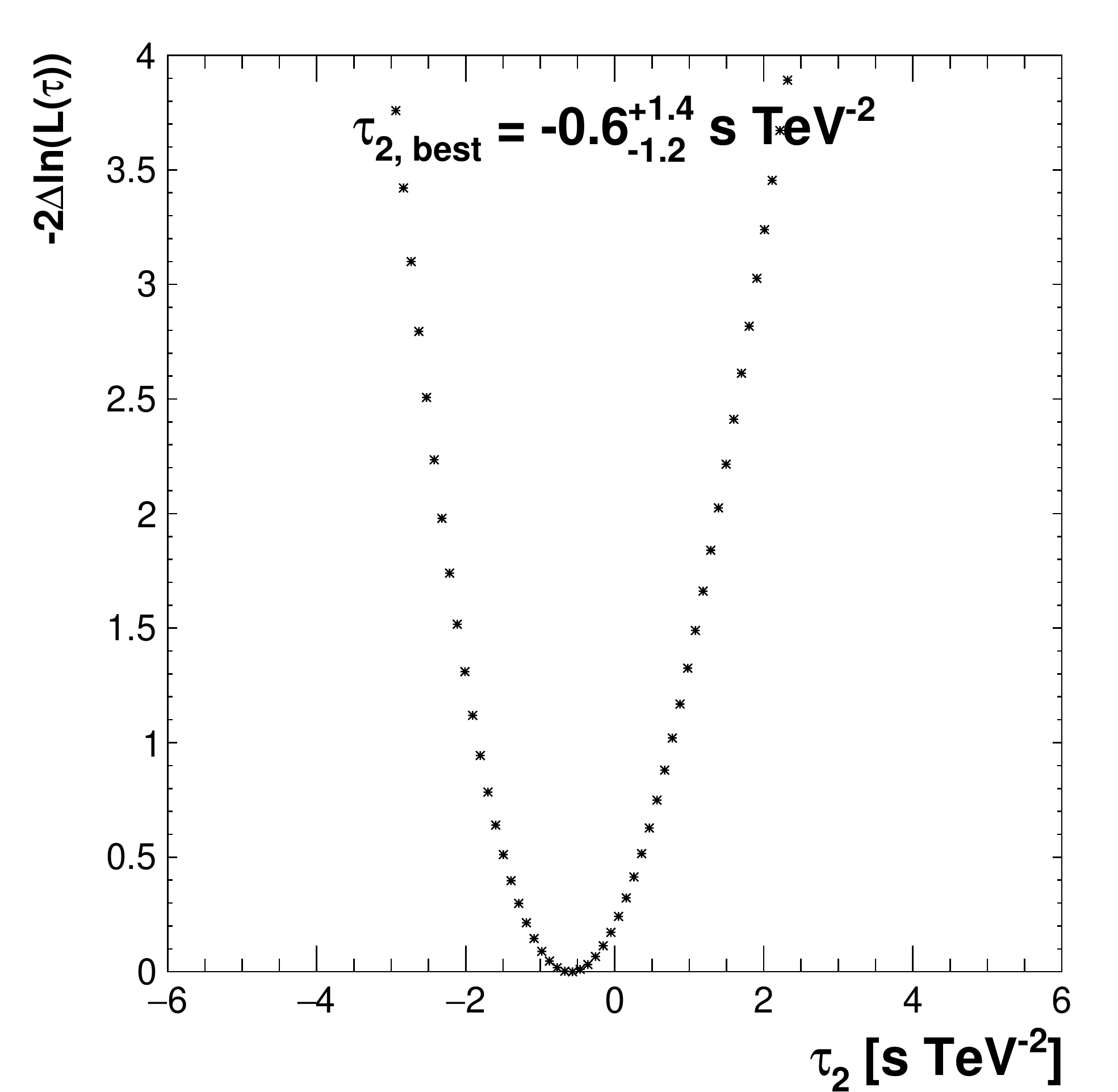}
\end{center}
\caption{Likelihood function obtained from \mrk data for linear (left) and quadratic (right) models. The best fit values $\tau_{n, \text{best}}$ are given with their 1$\sigma$ errors.}
\label{likelihood}
\end{figure*}

The $\tau_{n,\text{best}}$ value of the LIV estimator is defined as the $\tau_n$ value minimizing the $-2\ln(L)$  function. Fig.~\ref{likelihood} presents the log-likelihood functions for the linear (left) and the quadratic (right) models. Each curve has a quadratic behavior and shows a single minimum. No significant energy-dependent time-lag is measured.

\begin{table}
	\caption{Parameters of the function $F_s(t)$.}
	\begin{center}
		\begin{tabular}{c c c} \hline
			Parameters							& Value	&	Error\\ \hline \hline
			$A_1$ \footnote{\label{flux}Expressed in 10$^{-12}$ cm$^{-2}\cdot s^{-1}$.} 	& 80.5	& 	6	\\ 
			$\mu_1$ (s)							& 2361 	& 	185	\\ 
			$\sigma_1$ (s)				    & 2153	& 	301	\\
			\hline
			$A_2$ \footref{flux}		& 60.5 & 	11	\\ 
			$\mu_2$ (s) 							& 6564	& 	220	\\ 
			$\sigma_2$ (s)					& 676 	& 	283	\\ 
			\hline
		\end{tabular}
	\end{center}
	\label{temp_par}
\end{table}

The statistical uncertainties quoted on Fig.~\ref{likelihood} are derived by requesting $-2\Delta \log(L) = 1$. These values are obtained from one realization and may be over- or under-estimated. Calibrated statistical uncertainties are considered instead, as derived from the ML analysis of 1000 simulated data sets mimicking actual data, \textit{i.e.} with identical light curve and spectral shape and no LIV time-lag. The resulting distributions of reconstructed $\tau_{n,\text{best}}$ parameter for $n = 1, 2$ are normally distributed, and their standard deviations are considered as calibrated statistical errors.

Systematic uncertainties are also estimated using simulations by looking at the induced variations on the reconstructed $\tau_{n,\text{best}}$ distribution when the spectral index and $F_s(t)$ parameters are smeared within their error intervals and when changing energy intervals boundaries according to the energy resolution. The ML analysis is also applied to photon lists from cross-check analyses to check the influence of reconstruction methods on the measured lag. The most important sources of systematic uncertainties are found to be related to the determination of $F_s(t)$, mainly the position of the peaks as already pointed out in \citet{HESS:2011aa}, and to the analysis chain. A possible contribution of the background is also investigated and found to be negligible.

The obtained values of $\tau_{n, \text{best}}$, with their $1\sigma$ statistical and overall systematic errors are:

\begin{align*}
            \tau_{1,\text{best}} = -8.2 \pm 21.5_{(stat)} \pm 14.2_{(syst)}\ \mathrm{s \cdot TeV^{-1}},\\
            \tau_{2,\text{best}} = -0.6 \pm 1.8_{(stat)} \pm 0.7_{(syst)}\ \mathrm{s \cdot TeV^{-2}}.
\end{align*}

These values are subsequently used to compute the 95\% confidence level limits on the quantum gravity energy scale $E_{QG}$, following Eq.~\ref{eq:timez5}. For the subluminal and superluminal scenarios, the obtained limits are :

 \begin{align*}
            E_{QG,1} > \begin{cases}
            3.6 \times 10^{17}\ \mathrm{GeV} & \text{(subluminal),}\\
            2.6 \times 10^{17}\ \mathrm{GeV} & \text{(superluminal),}\\
            \end{cases}
            \\[10pt]
            E_{QG,2} > \begin{cases}
            8.5 \times 10^{10}\ \mathrm{GeV} & \text{(subluminal),}\\
            7.3 \times 10^{10}\ \mathrm{GeV} &\text{(superluminal).}\\
            \end{cases}                
\end{align*}

\section{Spectral study}
\label{Section:Spec}

The energy spectrum of \mrk is obtained using the forward folding method described in \citet{Piron:2001ir}. The energy threshold used in the spectral analysis is defined as the energy at which the effective area reaches $15\%$ of its maximum value, yielding a threshold of $1.3\,$TeV. The energy spectrum extends up to $\sim 20\,$TeV, as shown in Fig.~\ref{Fig:Spec}.
A simple power law shape does not provide a good fit to the data, as the observed spectrum is significantly curved. This curvature can be interpreted in terms of attenuation of the intrinsic spectrum on the extragalactic background light (EBL). 

\subsection{EBL absorption and \mrk flare spectrum}
\label{SubSec:Spectrum}
The EBL is the background photon field originating from the integrated starlight and its re-processing by dust over cosmic history. It covers wavelengths ranging from the ultraviolet to the far-infrared. 
VHE $\gamma$ rays traveling over cosmological distances can interact with EBL photons and produce electron-positron pairs ($\gamma \gamma \rightarrow  e^+e^-$), resulting in an attenuated observed VHE flux above the pair production threshold \citep{Nikishov62, Gould:1967zzb, Stecker:1992wi}. 
The observed VHE spectrum of a blazar $\Phi_{\mathrm{obs}}(E_\gamma)$ at a redshift $z_s$ is the product of its intrinsic spectrum $\Phi_{\mathrm{int}}(E_\gamma)$ with the EBL attenuation effect: 

\begin{equation}
\Phi_{\mathrm{obs}}(E_\gamma)= \Phi_{\mathrm{int}}(E_\gamma) \times e^{-\tau(E_\gamma,z_s)},
\label{Eq:Attenuation}
\end{equation}
where $\tau(E_\gamma,z_s)$\footnote{The letter $\tau$ is used here for consistency with established nomenclature although it has been previously used in a different context in the previous section.} is the optical depth to \gs rays of observed energy $E_\gamma$. It takes into account the density of EBL photons $n_{\mbox{\tiny{EBL}}}$ and consists in an integration over the redshift $z$, the energy of EBL photons $\epsilon$, and the angle between the photon momenta $\theta$:

\begin{align}
\tau(E_\gamma,z_s) =&  \int_0^{z_s} dz \frac{dl}{dz} \int_{\epsilon_{thr}}^\infty d \epsilon \frac{dn_{\mbox{\tiny{EBL}}}}{d\epsilon}(\epsilon,z) \int_0^2 d\mu \frac{\mu}{2} \ \sigma_{\gamma \gamma} \left(s \right),
\label{Eq:Tau}
\end{align}
where $\mu = 1-\cos(\theta)$, and $\sigma_{\gamma \gamma}$ is the pair production cross section \citep{Breit:1934zz}. The square of the center of mass energy $s$ for an interaction with a \gs ray of energy $E_\gamma' = (1+z) E_\gamma$ is given by

\begin{equation}
s = 2 E_\gamma' \epsilon \mu,
\label{Eq:s}
\end{equation}
and the threshold EBL photon energy for pair production  $\epsilon_{thr}$ in the case of a head-on collision ($\theta= \pi$) is

\begin{align}
\epsilon_{thr} (E_\gamma',z)=& \frac{m_e^2 c^4}{ E_{\gamma}' (1+z)}  \nonumber \\ &  \simeq \frac{0.26}{(1+z)} \left(  \frac{E_\gamma'}{\rm{TeV}}\right)^{-1} \rm{ eV}.
\label{Eq:Threshold}
\end{align}

EBL attenuation leaves a redshift- and energy-dependent imprint on the observed spectrum of blazars and can be used to probe the spectral energy distribution (SED) of the EBL. Knowledge of the EBL SED has greatly improved over the last decade. Predictions from models \citep{Franceschini:2008tp, Dominguez:2010bv, Finke:2009xi, Gilmore:2011ks}, constraints from $\gamma$ rays \citep{Meyer:2012us,Biteau:2015xpa,H.E.S.S.:2017odt}, and results from empirical determinations \citep{Stecker:2016fsg} agree between lower and upper limits. In the following, the model of \citet{Franceschini:2008tp} is used as a reference.

Despite a low redshift of $z=0.034$, EBL attenuation for Mrk~501 is non-negligible at energies larger than 1~TeV. The associated optical depth reaches 1 around 10~TeV \citep{Franceschini:2008tp},  corresponding to mid-infrared EBL wavelengths (Eq.~\ref{Eq:Threshold}).\\

\begin{figure}[]
\includegraphics[scale=0.44]{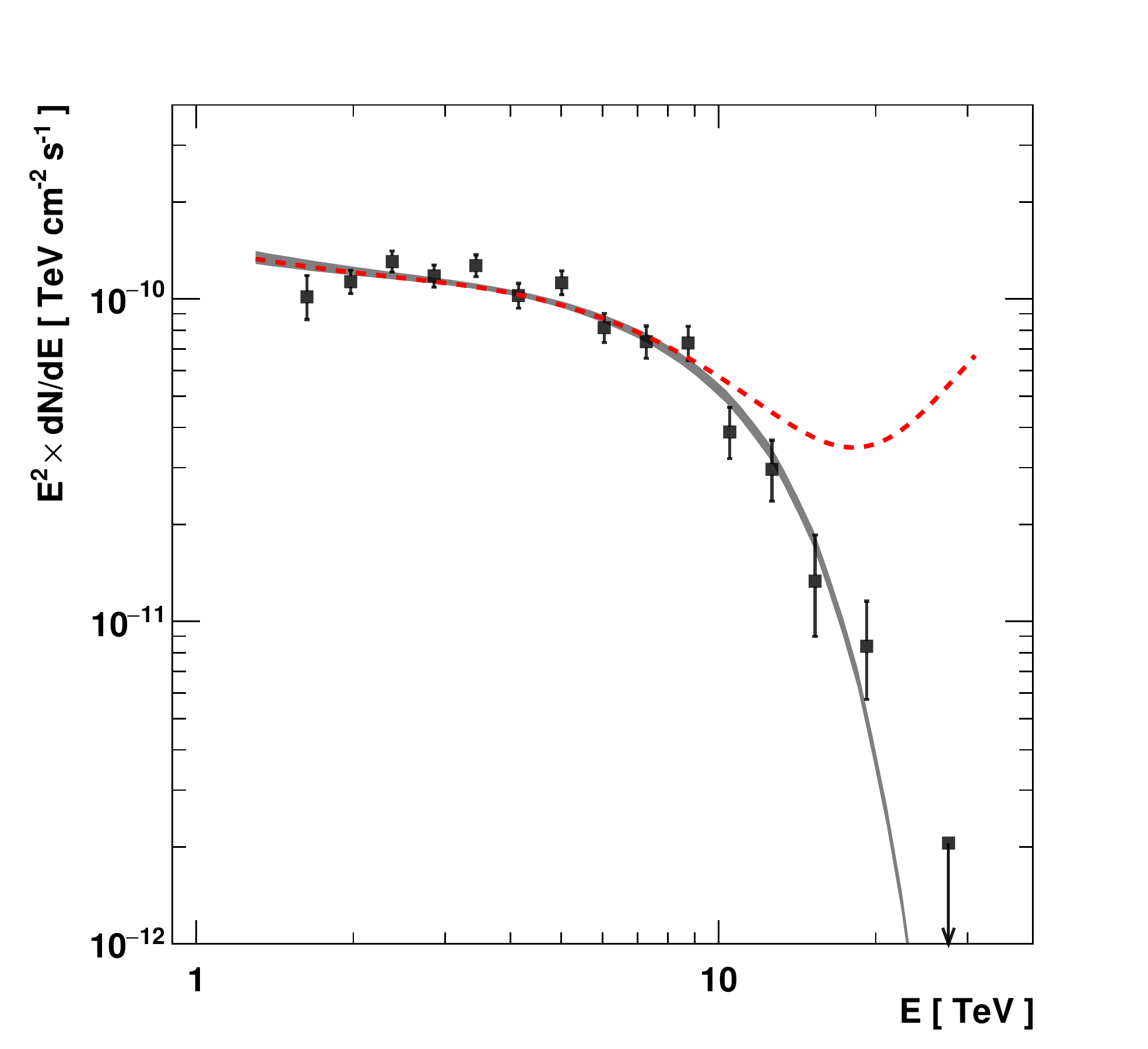}
\caption{Energy spectrum observed from the flare of Mrk~501. The best-fit EBL-attenuated power law is displayed by a solid line. The spectral points are obtained from residuals to the fit. A minimum significance of 3$\sigma$ is required for each point. The red dashed line represents the expected spectrum for the same intrinsic shape but considering subluminal linear LIV with $E_{\rm{QG},1}= E_{\rm{Planck}}$.}
\label{Fig:Spec}
\end{figure}

The \mrk flare intrinsic spectrum measured by \hess is well fitted by an intrinsic power law $(\Phi_{\mathrm{int}}(E_\gamma) = \phi_0 E_\gamma^{-\alpha}$) attenuated on the EBL using the optical depth of the model of \citet{Franceschini:2008tp}, as shown in Fig~\ref{Fig:Spec}. The fitted intrinsic index is $\alpha = 2.03 \pm 0.04_{\rm{stat}} \pm 0.2_{\rm{sys}}$.
Intrinsic shapes with curvature or a cut-off are not preferred over the simple power law. In this standard picture, EBL attenuation at the level of the model of \citet{Franceschini:2008tp} is sufficient to account for the entire observed curvature. The use of models with a significantly lower level of EBL density at infrared wavelengths would require intrinsic curvature. On the other hand, the use of models with a significantly higher level of EBL density at infrared wavelengths would cause an upturn in the intrinsic spectrum. This degeneracy is difficult to break, but current knowledge of the EBL SED gives good confidence that the VHE \mrk flare intrinsic spectrum follows a simple power law behavior up to $\sim$20~TeV.
The intrinsic power law shape is considered in the following as the natural choice accounting for the standard case. In the LIV case an intrinsic curvature could compensate for a genuine LIV effect. This degenerate scenario with no extrapolation to the standard case is not considered in this study.

\subsection{Opacity modifications due to LIV}
\label{SubSec:SpectralLIV}
The non-observation of deviations with respect to standard EBL attenuation at energies above 10~TeV can be used to put competitive constraints on $E_{\text{QG}}$. 
In the presence of LIV, the perturbation in the dispersion relation Eq.\ref{Eq:DispersionLIV} propagates into the EBL optical depth  (Eq.\ref{Eq:Tau}). The center-of-mass energy squared $s$ (Eq.~\ref{Eq:s}) and threshold energy for pair production $\epsilon_{\text{thr}}$ (Eq.~\ref{Eq:Threshold}) are modified with an extra term \citep{Tavecchio:2015rfa}:

\begin{align}
 s \rightarrow s \pm \frac{E_\gamma^{'n+2} }{E_{\text{QG}}^n} \text{,		 and  	  } \epsilon_{\text{thr}} \rightarrow \epsilon_{\text{thr}} \mp \frac{1}{4} \frac{E_\gamma^{'n+1} }{E_{\text{QG}}^n}. 
\end{align}

It is assumed that the modified center-of-mass energy squared  $s$ is still an invariant quantity in the LIV framework  \citep{Fairbairn:2014kda, Tavecchio:2015rfa}. The effects of LIV on electrons are neglected, as the constraints on the LIV scale for electrons are stringent \citep{Liberati_LIV_electrons}. 

In the context of investigations for a potential transparency excess of the Universe to VHE \gs rays  (as hinted at in \citealt{Horns:2012fx}), only the subluminal case (minus sign in Eq. \ref{Eq:DispersionLIV}) is considered: if non negligible, the LIV term will induce lower values for $s$ (higher threshold value $\epsilon_{\text{thr}}$) suppressing pair creation on the EBL, therefore causing an excess of transparency of the Universe to the most energetic $\gamma$ rays \footnote{An excess of transparency of the Universe to \gs rays could also be caused by the conversion of photons to axion-like particles in magnetic fields, see \textit{e.g.} \citet{2009PhRvD..79l3511S}.}. 

In the subluminal LIV scenario, the threshold energy is given by

\begin{equation}
 \epsilon_{\text{thr}} =  \frac{m_e^2 c^4}{E_\gamma'} +  \frac{1}{4} \frac{E_\gamma^{'n+1} }{E_{\text{QG}}^n}.
 \label{Eq:LIVThr}
\end{equation}

This threshold energy is no longer a monotonic function in $E_\gamma$. The critical \g-ray energy corresponding to the minimal threshold energy can be obtained from the derivative of Eq.~\ref{Eq:LIVThr}. For linear ($n=1$) perturbations, this critical energy is $18.5$~TeV $\left(\frac{E_{\rm{QG},1}}{E_{\rm{Planck}}} \right)^{1/3}$. Extragalactic \gs rays at this energy can thus probe Planck scale linear LIV\footnote{Planck scale is, however, out of reach in the case of quadratic ($n=2$) perturbations, as the critical energy in this case is $\sim~8~\times~10^{4}$~TeV $\left(\frac{E_{\rm{QG},2}}{E_{\rm{Planck}}} \right)^{1/2}$.}, as shown by the red dashed line on Fig.~\ref{Fig:Spec}. 

\subsection{Constraints on the LIV scale}
\begin{figure}
\begin{center}
\subfigure[\label{TS_n1}]{\includegraphics[scale=0.35]{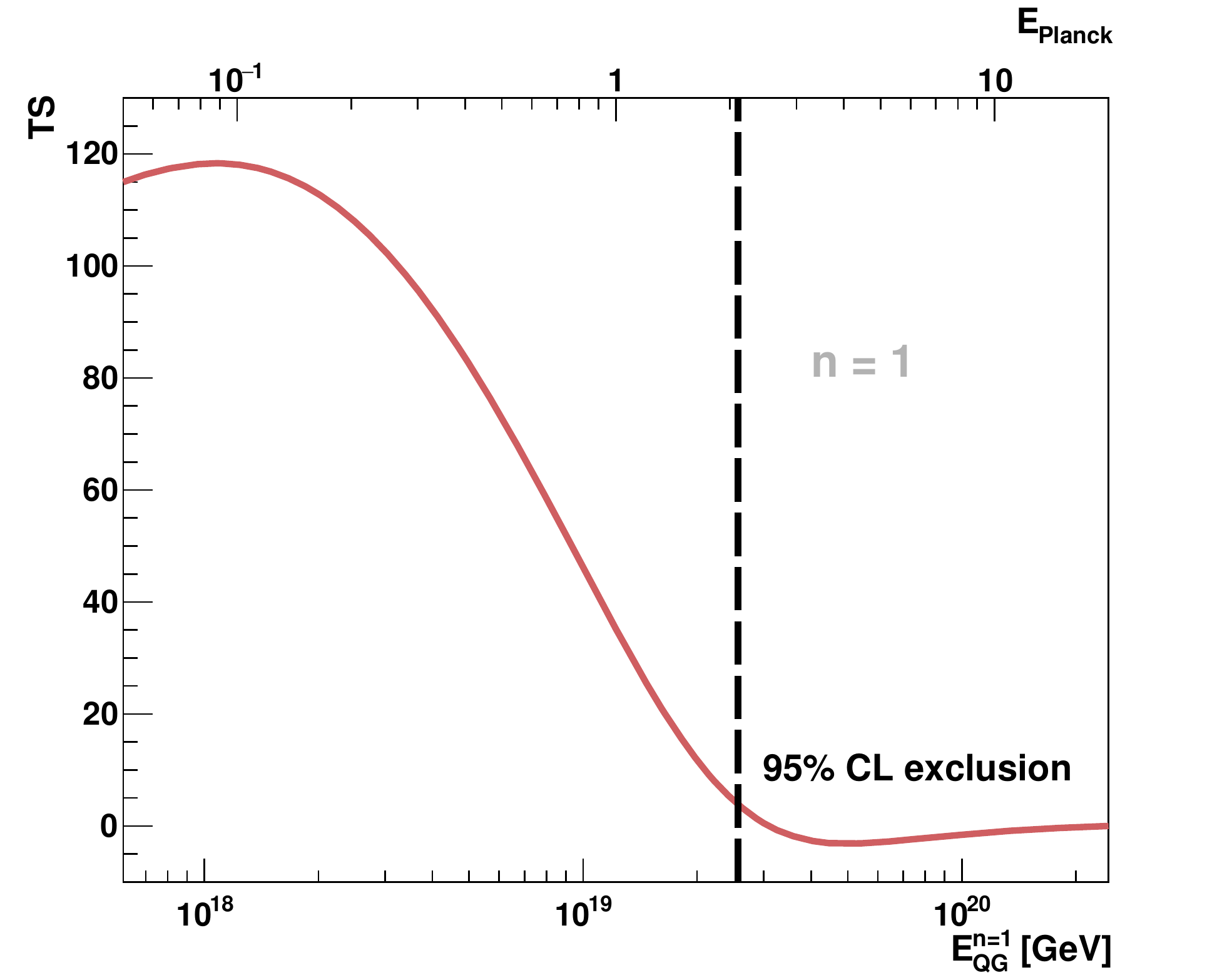}}
%\vspace*{-3mm}
\subfigure[\label{TS_n2}]{\includegraphics[scale=0.35]{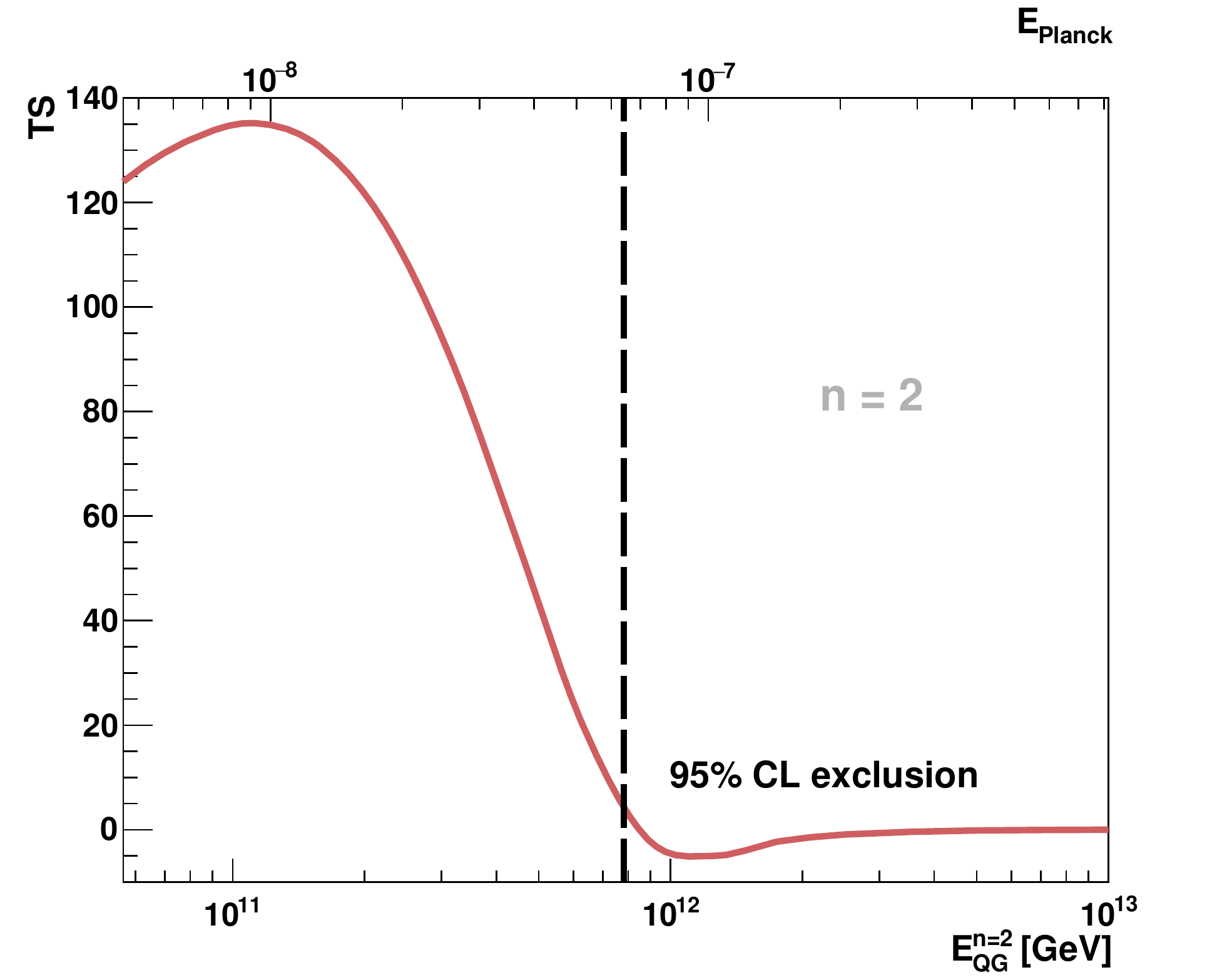}}
\end{center}

\caption{TS profiles obtained from the fit of the flare spectrum to an intrinsic power law absorbed on the EBL model of \citet{Franceschini:2008tp} for the case of subluminal linear (\ref{TS_n1}) and quadratic (\ref{TS_n2})  LIV perturbations. The black dashed line corresponds to the lower limit on $E_{\rm{QG}}$ at 95\% confidence level.}
\end{figure}

Optical depths to \gs rays using the EBL SED of the model of \citet{Franceschini:2008tp} are computed considering modifications due to subluminal LIV for linear and quadratic perturbations. The forward folding fit of the \mrk flare spectrum is performed assuming an intrinsic power law with spectral index and normalization free in the fit. Values of $E_{\rm{QG}}$ are scanned logarithmically in the range of interest for observable deviations in the covered energy range.  As the spectrum shows no evidence for an upturn, LIV-free optical depth values are preferred and the best fit $\chi^2$ values reach plateaus corresponding to the standard case. In order to quantify this effect, the following test statistic is considered: $\rm{TS} = \chi^2(E_{\rm{QG}}) - \chi^2(E_{\rm{QG}} \rightarrow  \infty)$, where $E_{\rm{QG}} \rightarrow  \infty$ corresponds to the standard case. TS profiles for linear and quadratic cases are represented in Fig.~\ref{TS_n1} and Fig.~\ref{TS_n2} respectively. 

From these TS profiles exclusion limits on $E_{\rm{QG}}$ are obtained. In the linear case the limit $E_{\rm{QG},1} >  2.6 \times 10^{19} \rm{GeV} \ (\textit{i.e. } 2.1 \times E_{\rm{Planck}})$ is obtained at 95\% confidence level. $E_{\rm{Planck}}$ is excluded at the 5.8$\sigma$ level. These results are comparable with the limits obtained using the 1997 flare spectrum of \mrk observed by HEGRA \citep{Biteau:2015xpa, Tavecchio:2015rfa}. These Planck scale limits on linear LIV are competitive with the best limits obtained considering time delays with GRBs. In the quadratic case, the limit $E_{\rm{QG},2} > 7.8 \times 10^{11} \rm{ GeV} \ (\textit{i.e. } 6.4 \times 10^{-8} \times E_{\rm{Planck}})$ is obtained at 95\% confidence level. This is the best existing limit on quadratic LIV perturbations to the dispersion relation of photons.

The main source of uncertainty on the derived limits on $E_{\rm{QG}}$ through this spectral method is the degeneracy between the spectral upturn caused by LIV and the possibility of an intrinsic upturn, together with the uncertainty related to EBL attenuation. Using the lower-limit EBL model of \citet{2010A&A...515A..19K} the value of $E_{\rm{QG}}$ required for an equivalent flux attenuation at 20~TeV would be six times higher than the value using the EBL model of \citet{Franceschini:2008tp}. The above limits are valid considering the natural interpretation that the intrinsic VHE spectrum of the \mrk flare has a power law behavior and is attenuated using state-of-the-art EBL models.

\section{Discussion and conclusions}
\label{Section:Summary}
The observation of a bright flare of \mrk with H.E.S.S. in June 2014 reveals multi-TeV variability on minutes timescales and an energy spectrum extending up to 20~TeV compatible with a simple power law attenuated by the EBL. These characteristics make this flare a unique opportunity to probe LIV in the photon sector with \hess using both temporal and spectral methods. Competitive results on the LIV energy scale $E_{\rm{QG}}$ are obtained considering linear or quadratic perturbations in the dispersion relation of photons.
Temporal and spectral methods are kept separate as a proper combination of results is considered complex due to the very different analysis procedures. Such a combination would moreover not be beneficial to the LIV constraints given the order of magnitude separating the results from both approaches.

Using the temporal method, the limit for the linear case considering a subluminal LIV effect is similar to the one obtained by H.E.S.S. using PG~1553+113 data \citep{Abramowski:2015}. For the quadratic case, the limit obtained is the best time-of-flight limit obtained with an AGN, slightly above the one obtained by H.E.S.S. with PKS 2155-304 \citep{HESS:2011aa}. This follows from the exceptional energy coverage of this flare with a substantial sample of photons above 10~TeV.

Assuming the EBL-attenuated power law spectral behavior presented in \ref{SubSec:Spectrum} and the framework described in \ref{SubSec:SpectralLIV}, the spectral method yields an exclusion limit for the linear case above the Planck energy scale and the best existing limit for the quadratic case. Thus it places the blazar flare studies  with VHE \g-ray astronomy instruments at the level of the time-of-flight limits obtained with GRBs (\textit{e.g.} GRB 090510 \citealt{Vasileiou:2015wja}). 

These results will be useful for LIV studies combining data from several \g-ray instruments as in \citet{Noguès2017}. This is particularly promising in the context of the advent of the CTA observatory \citep{Acharya:2017ttl} which will allow population studies with unprecedented sensitivity.

\acknowledgments
\footnotesize{
The support of the Namibian authorities and of the University of Namibia in facilitating 
the construction and operation of H.E.S.S. is gratefully acknowledged, as is the support 
by the German Ministry for Education and Research (BMBF), the Max Planck Society, the 
German Research Foundation (DFG), the Helmholtz Association, the Alexander von Humboldt Foundation, 
the French Ministry of Higher Education, Research and Innovation, the Centre National de la 
Recherche Scientifique (CNRS/IN2P3 and CNRS/INSU), the Commissariat \`a l'\'energie atomique 
et aux \'energies alternatives (CEA), the U.K. Science and Technology Facilities Council (STFC), 
the Knut and Alice Wallenberg Foundation, the National Science Centre, Poland grant no. 2016/22/M/ST9/00382, the South African Department of Science and Technology and National Research Foundation, the University of Namibia, the National Commission on Research, Science \& Technology of Namibia (NCRST), the Austrian Federal Ministry of Education, Science and Research and the Austrian Science Fund (FWF), the Australian Research Council (ARC), the Japan Society for the Promotion of Science and by the University of Amsterdam. We appreciate the excellent work of the technical support staff in Berlin, Zeuthen, Heidelberg, Palaiseau, Paris, Saclay, T\"ubingen and in Namibia in the construction and operation of the equipment. This work benefited from services provided by the H.E.S.S. 
Virtual Organisation, supported by the national resource providers of the EGI Federation.
This work benefits from the triggers received from the FACT collaboration.
}

\bibliography{FlarePaperBiblio-1}
\bibliographystyle{aasjournal}

\end{document}